\documentclass{llncs}
\usepackage{graphicx}
\usepackage{caption}
\usepackage{multirow}
\usepackage{booktabs}

\usepackage{xcolor}

\begin{document}

\title{Joint Motion Estimation and Segmentation from Undersampled Cardiac MR Image}

\author{Chen Qin\inst{1}, Wenjia Bai\inst{1}, Jo Schlemper\inst{1}, Steffen E. Petersen\inst{2}, Stefan K. Piechnik\inst{3}, Stefan Neubauer\inst{3}, and Daniel Rueckert\inst{1}
}

\institute{Department of Computing, Imperial College London, London, UK\\
\and
NIHR Biomedical Research Centre at Barts, Queen Mary University of London, London, UK\\
\and Division of Cardiovascular Medicine, Radcliffe Department of Medicine, University of Oxford, Oxford, UK\\}

\maketitle              

\begin{abstract}
Accelerating the acquisition of magnetic resonance imaging (MRI) is a challenging problem, and many works have been proposed to reconstruct images from undersampled $k$-space data. However, if the main purpose is to extract certain quantitative measures from the images, perfect reconstructions may not always be necessary as long as the images enable the means of extracting the clinically relevant measures. In this paper, we work on jointly predicting cardiac motion estimation and segmentation directly from undersampled data, which are two important steps in quantitatively assessing cardiac function and diagnosing cardiovascular diseases. In particular, a unified model consisting of both motion estimation branch and segmentation branch is learned by optimising the two tasks simultaneously. Additional corresponding fully-sampled images are incorporated into the network as a parallel sub-network to enhance and guide the learning during the training process. Experimental results using cardiac MR images from 220 subjects show that the proposed model is robust to undersampled data and is capable of predicting results that are close to that from fully-sampled ones, while bypassing the usual image reconstruction stage. 

\end{abstract}
\section{Introduction}
Cardiac magnetic resonance imaging (MRI) provides qualitative and quantitative information of the morphology and function of the heart, which are crucial for assessing cardiovascular diseases. Both cardiac MR image segmentation and motion estimation are essential steps for the dynamic exploration of the cardiac function.
However, one limitation of the cardiovascular MR is the low acquisition speed due to both hardware and physiological constraints. Most approaches consider undersampling the data in $k$-space and then reconstruct the images \cite{qin2017convolutional,schlemper2017dynamic}. Nevertheless, in most cases, perfect reconstructions are not necessary as long as the images allow to obtain accurate clinically relevant parameters such as changes in ventricular volumes and the elasticity and contractility properties of the myocardium. Therefore, instead of firstly recovering non-aliased images, it may be more effective to estimate the final results directly from undersampled MR data and also to make such estimations as accurate as possible.

In this paper, we propose to learn a joint deep learning network for cardiac motion estimation and segmentation directly from undersampled cardiac MR data, bypassing the MR reconstruction process. In particular, we extend the joint model proposed in \cite{qin2018} which consists of an unsupervised cardiac motion estimation branch and a weakly-supervised segmentation branch, where the two tasks share the same feature encoder. We investigate the network's capability of predicting motion estimation and segmentation maps simultaneously and directly from undersampled cardiac MR data. The problem is formulated by incorporating supervision from fully sampled MR image pairs in addition to the composite loss function as proposed in \cite{qin2018}. Simulation experiments have been performed on 220 subjects under different acceleration factors with radial undersampling patterns. Experiments indicate that results learned directly from undersampled data are reasonably accurate and are close to predictions from fully-sampled data. This could potentially lead to future works that enable fast and accurate analysis in an integrated MRI reconstruction and analysis pipeline.

\subsection{Related Work}
Cardiac segmentation and motion estimation are well studied problems in medical imaging. Traditionally, most approaches consider these two tasks separately \cite{bai2017human,shi2012comprehensive,tobon2013benchmarking}. However, it is known that segmentation and motion estimation problems are closely related, and optimising these two tasks jointly has been proven to improve the performance for both challenges. Recently, Oksuz et al. \cite{oksuz2017joint} proposed a joint optimisation scheme for registration and segmentation using dictionary learning based descriptors, which enables better performance for both of these ill-posed processes. Qin et al. \cite{qin2018} proposed a unified deep learning model for both cardiac motion estimation and segmentation, where no motion ground truth is required and only temporally sparse annotated frames in a cardiac cycle are needed.


However, there are only a limited number of works that focus on obtaining segmentation maps and motion fields directly from undersampled MR data. One direction of the research is on the application-driven MRI \cite{caballero2014application}, where an integrated acquisition-reconstruction-segmentation process was adopted to provide a more efficient and accurate solution. Schlemper et al. \cite{jo2018} expanded on the idea of application-driven MRI and presented an end-to-end synthesis network and a latent feature interpolation network to predict segmentation maps from extremely undersampled dynamic MR data. Our work focuses on the scenario where motion fields and segmentation maps can be jointly predicted directly from undersampled MR data, bypassing the usual MR image reconstruction stage.

\section{Methods}
\begin{figure*}[!t]
\centering
\includegraphics[width=\linewidth]{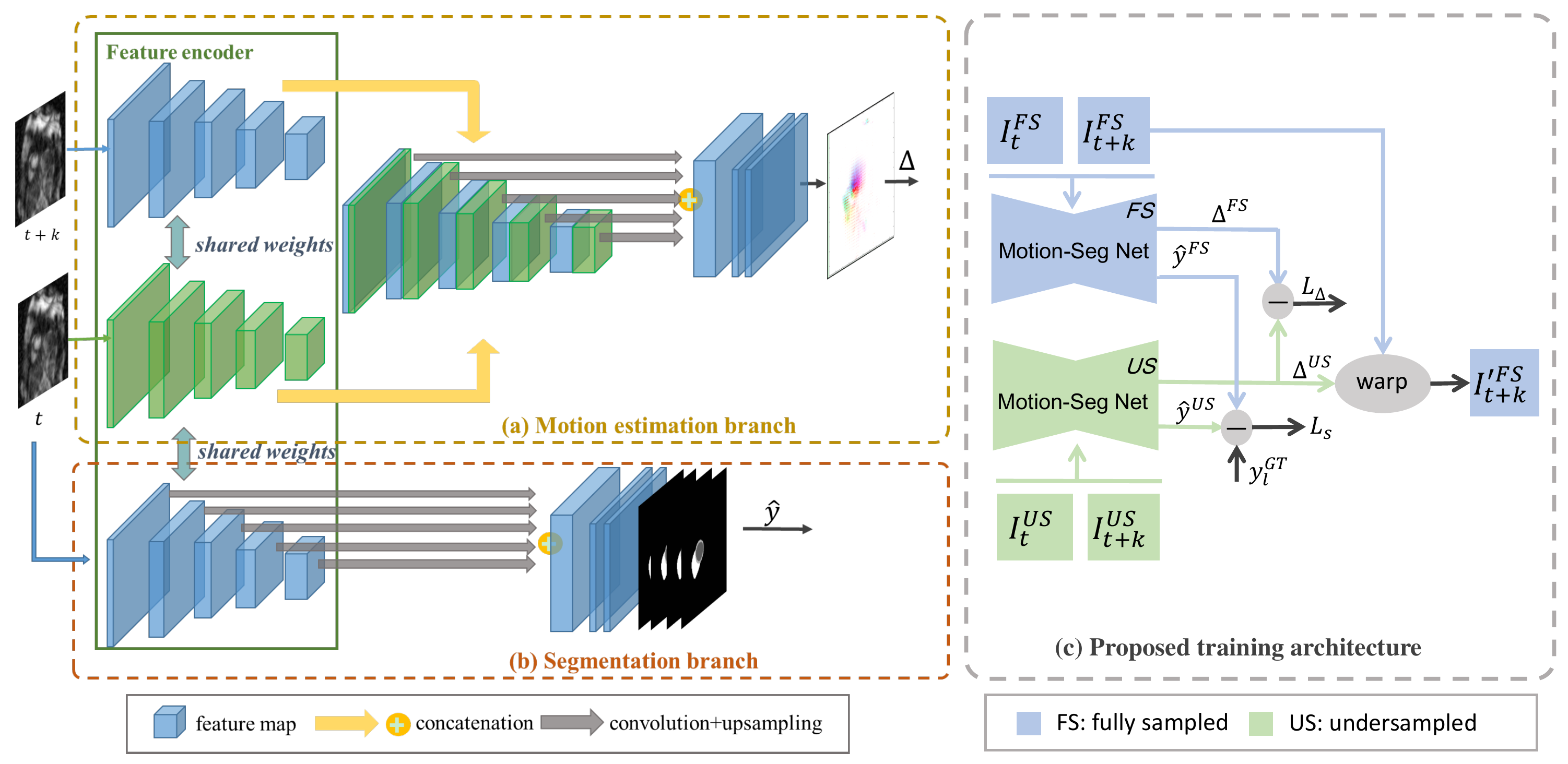}
\caption{The overall schematic architecture of proposed network for joint estimation of cardiac motion and segmentation directly from undersampled data. (a) (b) The Motion-Seg net adopted from \cite{qin2018}. (c) Proposed architecture for training the Motion-Seg net on undersampled data. US: undersampled, FS: fully-sampled.}
\label{fig:joint_model}
\end{figure*}

Our goal is to predict the simultaneous motion estimation and segmentation directly from undersampled cardiac MR images and make sure that such predictions are as accurate and efficient as possible. Here we extend the effective unified model (Motion-Seg Net) proposed in \cite{qin2018} to adapt to the application for undersampled MR data. The proposed network architecture consists of two branches which perform motion estimation and segmentation jointly, and a well-trained sub-network for fully-sampled images is incorporated to provide additional supervision during the training process. Note that at test stage, only the undersampled sub-network is needed, and no fully-sampled data is required. The overall architecture of the model is shown in Fig. \ref{fig:joint_model}.
\subsection{Unsupervised Cardiac Motion Estimation from Undersampled MR Image}
\label{motion estimation}
Inspired by the success of the joint prediction network proposed in \cite{qin2018} which effectively learns useful representations, here we propose to adapt the network to undersampled MR data. In contrast to the fully-sampled case where only self-supervision is required for the motion estimation, it is difficult for the undersampled images to merely rely on self-supervision, i.e., the intensity difference, due to the noises caused by aliased patterns. To address this, we propose to incorporate their corresponding fully-sampled image pairs as an additional supervision to guide the training for the undersampled images, and a schematic illustration of the model is shown in Fig. \ref{fig:joint_model}(a)(c). 

The task is to find an optical flow representation between the target undersampled frame $I_{t}^{US}$ and the source undersampled frame $I_{t+k}^{US}$, where the output is a pixel-wise 2D motion field $\Delta^{US}$ representing the displacement in $x$ and $y$ directions. We exploit a modified version of the network proposed in \cite{qin2018} for the representation learning, in which it mainly consists of three components: a Siamese network for the feature extraction of both target frame and source frame where the encoder is adapted from VGG-16 Net; a multi-scale concatenation of features from pairs of frames motivated by the traditional multi-level registration method \cite{rueckert1999nonrigid}; and a bilinear interpolation sampler that warps the source frame to the target one by using the estimated displacement field ${\Delta}^{US} = ({\Delta}^{US} x, {\Delta}^{US} y; \theta_{\Delta}^{US})$, where the network is parameterised by $\theta_{\Delta}^{US}$ which is learned directly from undersampled MR data. Note that a RNN unit could be potentially incorporated to propagate motion information along the temporal dimension \cite{qin2018}, and we will leave it as one of our future work. 


Due to the severe aliased patterns existing in the undersampled MR images, it is not practical to train the spatial transformer network purely based on minimising the intensity difference between the transformed undersampled frame and the target undersampled frame. To address this, we propose to introduce the fully-sampled image pairs as a supervision for the training. Specifically, instead of warping the undersampled source image, here we propose to transform the corresponding fully-sampled source image, which can be expressed as $I_{t+k}^{'FS}(x,y) = \Gamma\{I_{t+k}^{FS}(x+\Delta_{t+k}^{US}x, y+\Delta_{t+k}^{US}y)\}$. Then the network can be trained by optimising the pixel-wise mean squared error between $I_{t}^{FS}$ and $I_{t+k}^{'FS}$. To ensure local smoothness, we maintain the regularisation term for the gradients of displacement fields which uses an approximation of Huber loss proposed in \cite{caballero2017real,qin2018}, namely $\mathcal{H}(\delta_{x,y}\Delta^{US}) = \sqrt{\epsilon+\sum_{i=x,y}(\delta_{x}\Delta^{US} i^2+\delta_{y}\Delta^{US} i^2)}$, where $\epsilon=0.01$.  Therefore, the loss function can be described as follows:
\begin{equation}
\label{loss_1}
\mathcal{L}_m = \frac{1}{N_s}\sum_{\left( I_t,I_{t+k} \right) \in S}\big[\|I_t^{FS}-I_{t+k}^{'FS}\|^2+\alpha \mathcal{H}(\delta_{x,y}\Delta_{t+k}^{US})\big],
\end{equation}
where $N_s$ stands for the number of sample pairs in the training set $S$, and $\alpha$ is a regularisation parameter to trade off between image dissimilarity and local smoothness.

However, it is observed that for heavily undersampled images, such weak supervision in Eq. \ref{loss_1} is not sufficient. Therefore, in order to push the learning results from undersampled data to be as accurate as that from fully-sampled data, we additionally introduce a pixel-wise mean squared error loss on the displacement fields between the estimation from undersampled data ($\Delta_{t+k}^{US}$) and that from fully-sampled one ($\Delta_{t+k}^{FS}$). Since only the motion of anatomical structures is of interest, here we propose to mask the region of interests (ROI) by utilising the predicted segmentation maps from fully-sampled data to allow that only errors from ROI can be backpropagated to contribute to the learning. The proposed loss term can be expressed as $\mathcal{L}_{\Delta_{t+k}}=\|(\Delta_{t+k}^{US}-\Delta_{t+k}^{FS}) * {\bf{M}}_t\|^2$, where ${\bf{M}}_t$ is a one-hot mask (1 for ROI, and 0 for background) generated from the segmentation maps from frame $t$ of fully-sampled images. Thus, the overall loss function for motion estimation is as follows: 
\begin{equation}
\label{loss_2}
\mathcal{L}_m = \frac{1}{N_s}\sum\big[\|I_t^{FS}-I_{t+k}^{'FS}\|^2+\alpha \mathcal{H}(\delta_{x,y}\Delta_{t+k}^{US}) + \beta \|(\Delta_{t+k}^{US}-\Delta_{t+k}^{FS})* {\bf{M}}_t\|^2  \big],
\end{equation}
in which an additional trade-off parameter $\beta$ is introduced. Note that no ground truth displacement fields are required during the training, thus the motion is still estimated unsupervisedly.

\subsection{Joint Cardiac Motion Estimation and Segmentation from Undersampled MR Image}
\label{joint model}
Previous works have shown that motion estimation and segmentation tasks are complementary \cite{cheng2017segflow,qin2018,tsai2016video}. Therefore, here we couples both tasks for the joint prediction from undersampled MR data. The schematic architecture of the unified model is shown in Fig. \ref{fig:joint_model}.

The joint model consists of two branches: the motion estimation branch proposed in Section \ref{motion estimation} which introduces additional supervision from fully sampled images, and the segmentation branch based on the network proposed in \cite{bai2017human}, where both branches share the joint feature encoder (Siamese style network) as shown in Fig. \ref{fig:joint_model}. As images are only temporally sparse annotated, predictions from corresponding fully-sampled images are used as supervision for those unlabelled data.
Therefore a categorical cross-entropy loss $\mathcal{L}_s = -\sum_{l\in L} y_{l}^{GT} \textup{log}( f(x_{l};\Theta^{US})) -\sum_{n\in U} \hat{y}_{n}^{FS} \textup{log}( f(x_{n};\Theta^{US}) )$ on labelled data set $L$ and unlabelled data set $U$ is used for segmentation branch, in which we define $x_l$ and $x_n$ as the input data, $y_l^{GT}$ as the ground truth, $\hat{y}_n^{FS}$ is predictions from fully-sampled images and $f$ is the segmentation function parameterised by $\Theta^{US}$.   
Different from the loss function as stated in \cite{qin2018}, here we don't employ the loss $\mathcal{L}_w$ between the warped segmentations and the target, as we find that for undersampled cases, minimising $\mathcal{L}_w$ could introduce more noises and uncertainties into the network training presumably because of the less accurate predictions. We empirically observed that this could lead to a small performance degradation especially for the segmentation branch.


As a result, the overall loss function for the joint model can be defined as:
\begin{equation}
\mathcal{L} = \mathcal{L}_{m}+\lambda\mathcal{L}_{s},
\end{equation}
where $\lambda$ is a trade-off parameter for balancing these two tasks. $\mathcal{L}_{m}$ can be of the form of Eq. \ref{loss_1} or Eq. \ref{loss_2}, and we will examine their comparisons in experiments.


\section{Experiments and Results}
Experiments were performed on 220 short-axis cardiac MR sequences from UK Biobank study. Each scan contains a sequence of 50 frames, where manual segmentations of left-ventricular (LV) cavity, the myocardium (Myo) and the right-ventricular (RV) cavity are available on ED and ES frames. A short-axis image stack typically consists of 10 image slices, and the pixel resolution is $1.8 \times 1.8 \times 10.0$ $mm^3$. Since only magnitude images are available, here we employed a phase map synthesis scheme proposed in \cite{jo2018} to synthetically generate phase maps (smoothly varying 2D sinusoid waves), in order to convert magnitude images to complex valued images and to make the simulation more realistic. In experiments, the synthesised complex valued images were back-transformed to regenerate k-space samples. The input undersampled images were generated by randomly undersampling the k-space samples using uniform radial undersampling patterns. For pre-processing, all training images were cropped to the same size of $192\times192$,  and intensity was normalized to the range of [0,1]. In our experiments, we split the data into 100/100/20 for training/testing/validation. Parameters used in the loss function were set to be $\alpha = 0.001$, $\beta = 1$, and $\lambda=0.01$, which were chosen via validation set. Fully-sampled sub-network parameters were loaded from \cite{qin2018}, and we train the undersampled network using Adam optimiser with a learning rate of 0.0001. Data augmentation was performed on-the-fly, with random rotation, translation, and scaling.

As work \cite{qin2018} has already shown that the joint model can significantly outperform model with single branch, in this work, we mainly focus on the evaluation of the performance on undersampled data. We first evaluated the performance of motion estimation by comparing the proposed model with a B-spline free-form deformation (FFD) algorithm\footnote{https://github.com/BioMedIA/MIRTK} \cite{rueckert1999nonrigid}, and the results are shown in Table 1. Here we examined the effect of different losses on the model's performance, where we termed method using $\mathcal{L}_m$ with the form of Eq. \ref{loss_1} as Proposed-A, and the one using Eq. \ref{loss_2} as Proposed-B. Motion fields were estimated between ES and ED frame, and mean contour distance (MCD) and Hausdorff distance (HD) were computed between the warped ES segmentations and ED segmentations. Results on fully-sampled (FS) images are presented in Table 1 as a reference. It can be observed that proposed methods consistently outperform FFD on all acceleration rates with $p \ll 0.001$ using Wilcoxon signed rank test, and is able to produce results that are close to the fully-sampled images. Furthermore, it can also be noticed that for higher acceleration rates ($6 \times$ and $8 \times$), Proposed-B produces significantly better results than Proposed-A ($p \ll 0.001$). This is reflected by the fact that higher undersampling rates result in more aliased images, therefore a relatively strong supervision ($\mathcal{L}_\Delta$) is more needed to guide the learning in comparison to images with less aliasing ($3 \times$).

\begin{table*}[!t]
  \centering
  \caption{Evaluation of motion estimation accuracy for undersampled MR data with different acceleration factors in terms of the mean contour distance (MCD) and Hausdorff distance (HD) in mm (mean and standard deviation). Loss function using $\mathcal{L}_m$(Eq. \ref{loss_1}) is termed as Proposed-A, and the one using $\mathcal{L}_m$(Eq. \ref{loss_2}) is termed as Proposed-B. Bold numbers indicate the best results for different undersampling rates.}
  \label{motion_evaluation}
\scalebox{0.9}
 { \begin{tabular}{cccccccc}
  
    \toprule
\multicolumn{2}{c}{\multirow{2}*{Method}} & \multicolumn{3}{c}{MCD} & \multicolumn{3}{c}{HD}\\
\cline{3-8}
 & & LV & {Myo}   & RV  & LV & {Myo}   & RV     \\
  \midrule
\multirow{2}{0.5cm}{FS} & {FFD} & 1.83 (0.53) & 2.47 (0.74) & 3.53 (1.25) & 5.10 (1.28)& 6.47 (1.69) & 12.04 (4.85)\\
& {Joint Model \cite{qin2018}} & \textbf{1.30} (\textbf{0.34})& \textbf{1.19} (\textbf{0.26}) & \textbf{3.03} (\textbf{1.08}) & \textbf{3.52} (\textbf{0.82}) & \textbf{3.43} (\textbf{0.87}) & \textbf{11.38} (\textbf{4.34})\\
   \midrule
\multirow{3}{0.5cm}{$3 \times$} & {FFD} & 2.19 (0.49) & 2.54 (0.74) & 3.94 (1.38) & 6.27 (1.64) & 6.62 (1.72) & 13.92 (5.03) \\
& {Proposed-A} & \textbf{1.32} (\textbf{0.40}) & \textbf{1.23} (\textbf{0.31}) & \textbf{3.41} (\textbf{1.22}) & \textbf{3.53} (\textbf{0.89}) & {3.59} ({1.10}) & \textbf{12.69} (4.47)\\
 & Proposed-B & 1.37 (0.45) & \textbf{1.23} (\textbf{0.31}) & 3.44 (1.22) & 3.59 (0.98) & \textbf{3.55} (\textbf{1.10}) & \textbf{12.69} (\textbf{4.45})\\
   \midrule
\multirow{3}{0.5cm}{$6 \times$} & {FFD} &2.80 (0.77) & 2.74 (0.75) & 4.48 (1.46)& 7.83 (2.30) & 7.26 (2.26) & 15.63 (5.19)\\
& {Proposed-A} & 2.10 (0.80)& 1.44 (0.38) & 3.84 (1.27) &4.79 (1.40)  & 3.98 (1.26) & 13.45 (4.49)\\
 & Proposed-B & \textbf{1.74} (\textbf{0.68})& \textbf{1.34} (\textbf{0.35}) & \textbf{3.68} (\textbf{1.27}) & \textbf{4.20} (\textbf{1.30}) & \textbf{3.77} (\textbf{1.21}) & \textbf{13.08} (\textbf{4.49})\\
   \midrule
\multirow{3}{0.5cm}{$8 \times$} & {FFD} & 3.29 (0.97) & 3.09 (0.99) & 4.94 (1.67) & 9.40 (2.70)& 8.48 (3.05) & 17.16 (5.75) \\
& {Proposed-A} & 2.30 (0.97)& 1.52 (0.46) & 4.02 (1.37) & 5.19 (1.71)& 4.16 (1.32) & 13.79 (4.60)\\
 & Proposed-B & \textbf{1.79} (\textbf{0.70}) & \textbf{1.44} (\textbf{0.39}) & \textbf{3.76} (\textbf{1.30}) & \textbf{4.36} (\textbf{1.40}) & \textbf{3.97} (\textbf{1.28}) & \textbf{13.27} (\textbf{4.55})\\
    \bottomrule
  \end{tabular}}
\end{table*}

\begin{figure*}[!t]
\centering
\includegraphics[width=\linewidth]{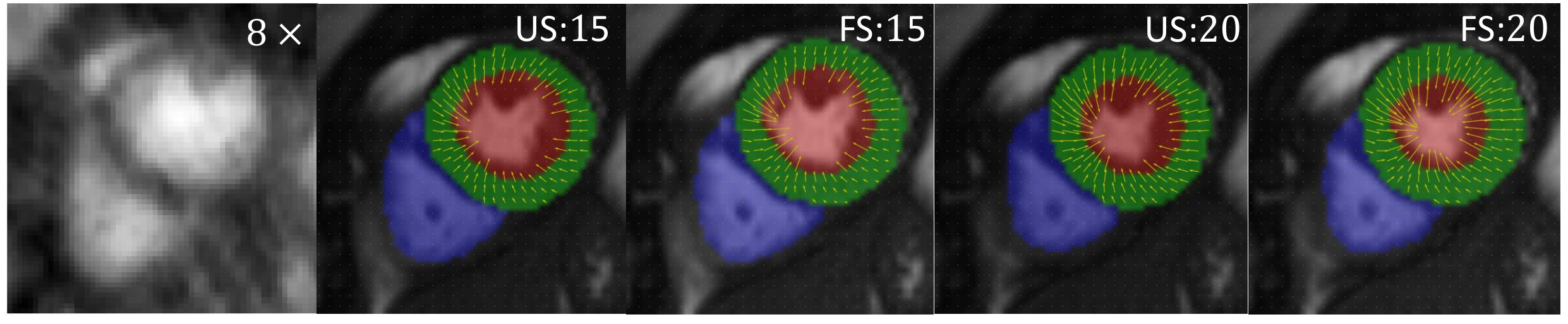}
\caption{Comparison visualisation results for simultaneous prediction of motion estimation and segmentation on data with undersampling rates 8. Myocardial motions are from ED to other time points (numbers on the top right). Segmentations are overlaid on fully-sampled data for better visualisation.}
\label{fig:visualization}
\end{figure*}

We further evaluated the segmentation performance of the model on undersampled data with different acceleration factors. Results reported in Table \ref{segmentaton_evaluation} are Dice scores computed with manual annotations on LV, Myo, and RV, as well as the clinical parameter ejection fraction (EF). It has been observed that Proposed-A and Proposed-B didn't differ significantly in terms of segmentation performance, so here we only report results obtained from Proposed-B in Table \ref{segmentaton_evaluation}. 
It can be seen that though there is a relatively small drop of performance as acceleration factors increase, the network is robust to train on undersampled data, and the clinical parameter predicted directly from undersampled data is very close to that from fully-sampled images. Furthermore, a visualisation result of the network predictions on $8 \times$ accelerated data in a cardiac cycle is shown in Fig. \ref{fig:visualization}, where myocardial motion indicated by the yellow arrows were established between ED and other time frames. Overall, predictions directly from undersampled MR data are reasonably accurate, despite some small underestimations.

\begin{table*}[!t]
  \centering
  \caption{Evaluation of segmentation performance under different acceleration factors in terms of Dice Metric (mean and standard deviation) and average percentage (\%) error for ejection fraction (EF) compared with fully-sampled data.}
  \label{segmentaton_evaluation}
   \setlength{\tabcolsep}{5.5pt}
  \begin{tabular}{ccccc}  
    \toprule
{Acceleration} & {LV} & {Myo}   & RV  & EF   \\
  \midrule
{FS} \cite{qin2018} &  \textbf{0.9348 (0.0408)} & \textbf{0.8640 (0.0295)} & {0.8861 (0.0453)} & -  \\
  {$3 \times$} & {0.9303 (0.0450)} & {0.8596 (0.0309)} & \textbf{0.8884 (0.0433)} & 2.68\%  \\
  {$6 \times$} & {0.9214 (0.0475)} & {0.8424 (0.0310)} & {0.8804 (0.0456)} & 3.56\%  \\
  {$8 \times$}  & {0.9141 (0.0487)} & {0.8260 (0.0343)} & {0.8658 (0.0523)} & 4.16\%  \\
    \bottomrule
  \end{tabular}
\end{table*}


\section{Conclusion}
In this paper, we explored the joint motion estimation and segmentation directly from undersampled cardiac MR data, bypassing the usual image reconstruction stage. The proposed method takes advantage of a unified model which shares the same feature encoder for both tasks and performs them simultaneously. In particular, we additionally introduced a parallel well-trained sub-network for corresponding fully-sampled MR image pairs as a supervision source for training undersampled data, in order to push the predictions from undersampled data to be as accurate as possible. We showed that the proposed network is robust to undersampled data, and results predicted directly from undersampled images are close to that from fully-sampled ones, which could potentially enable fast analysis for MR imaging. In the future, it is also interesting to explore methods that are independent of aliased patterns and acceleration factors. 

\bibliographystyle{splncs03}
\bibliography{ref}

\end{document}